\documentclass[epjST]{svjour}
\usepackage{graphicx,macros}
\usepackage{mathbbol}
\begin{document}
\title{Form factors in lattice QCD}
\author{
  B.B.~Brandt\inst{1} \and S.~Capitani\inst{1} \and M.~Della Morte\inst{1,2}
  \and D.~Djukanovic\inst{2} \and J.~Gegelia\inst{1} \and
  G.~von~Hippel\inst{1} \and A.~J\"uttner\inst{3} \and
  B.~Knippschild\inst{1} \and H.B.~Meyer\inst{1} \and
  H.~Wittig\inst{1,2}\fnmsep\thanks{\email{wittig@kph.uni-mainz.de}}}  
\institute{Institut f\"ur Kernphysik, University of Mainz, Becher Weg 45,
  D-55099 Mainz \and Helmholtz Institute Mainz, University of Mainz, D-55099
  Mainz \and CERN, Physics Department, TH Unit, CH-1211 Geneva 23}

\abstract{Lattice simulations of QCD have produced precise estimates
  for the masses of the lowest-lying hadrons which show excellent
  agreement with experiment. By contrast, lattice results for the
  vector and axial vector form factors of the nucleon show significant
  deviations from their experimental determination. We present results
  from our ongoing project to compute a variety of form factors with
  control over all systematic uncertainties. In the case of the pion
  electromagnetic form factor we employ partially twisted boundary
  conditions to extract the pion charge radius directly from the
  linear slope of the form factor near vanishing momentum transfer. In
  the nucleon sector we focus specifically on the possible
  contamination from contributions of higher excited states. We argue
  that summed correlation functions offer the possibility of
  eliminating this source of systematic error. As an illustration of
  the method we discuss our results for the axial charge, $\gA$, of
  the nucleon.}
%
\maketitle

\section{Introduction \label{s1int}}

Simulations of QCD on a space-time lattice have recently succeeded in
producing reliable results for several phenomenologically important
quantities. This enormous progress resulted mainly from improvements
of numerical techniques, which led to a significant acceleration of
simulation algorithms. A variety of hadronic observables can now be
computed with fully controlled statistical and systematic errors. A
well-known example is the spectrum of the lowest-lying mesons and
baryons. Following years of efforts by many different collaborations
\cite{qspect:CPPACS,qspect:CPPACS2,qspect:MILC98,qspect:ukqcd99,qspect:BGR03,dspect:CPPACS99,dspect:MILC01,dspect:PACS-CS08,dspect:BMW08,dspect:ETM0809},
it has now been firmly established that QCD accounts for the
experimentally observed spectrum within the quoted uncertainties of
the lattice calculation, which are at the level of a few percent. This
underlines once more that QCD is the correct theory of the strong
interaction, also in the low-energy regime.

\begin{figure}
\begin{center}
\includegraphics[width=9cm]{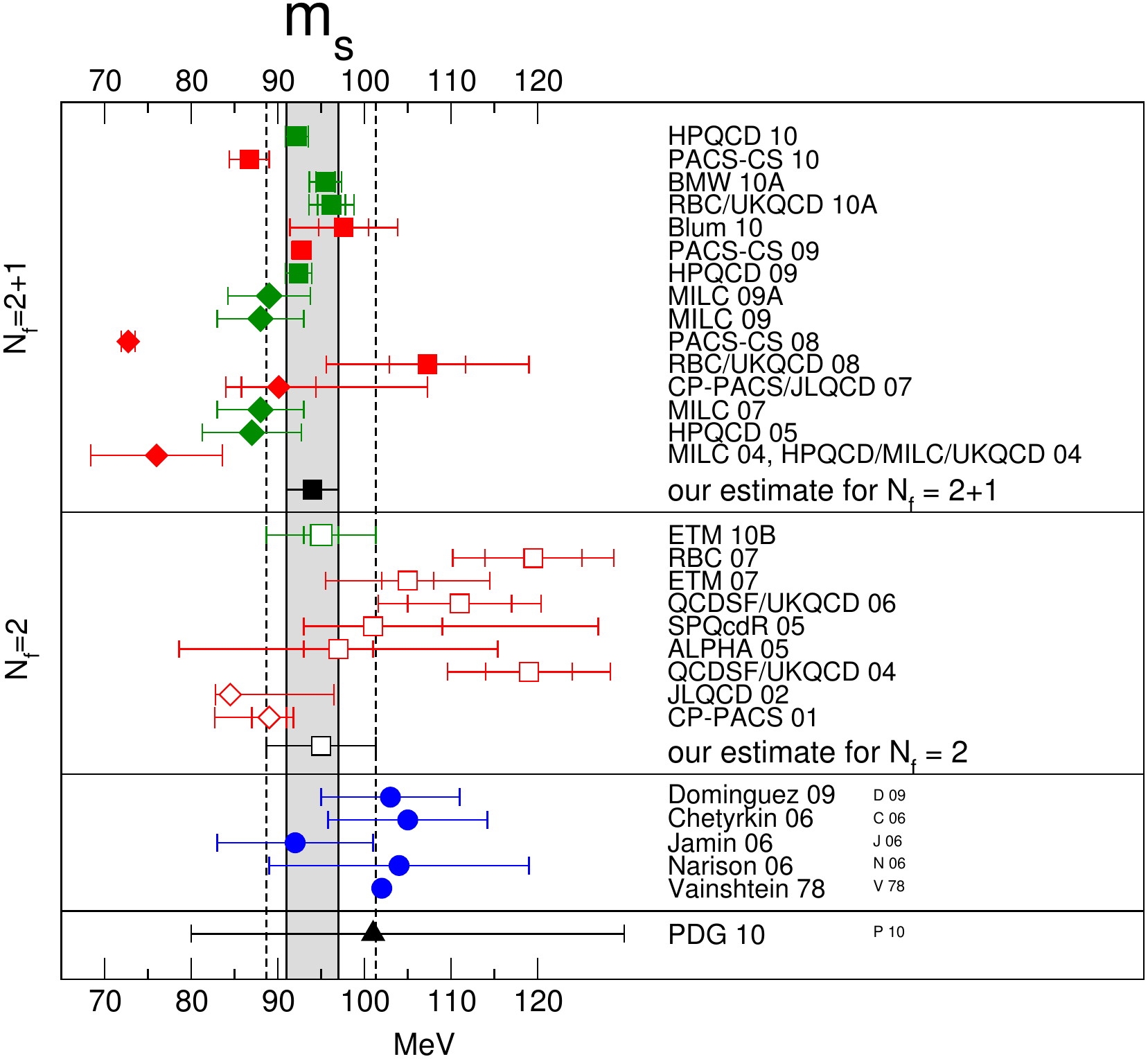}
\caption{Compilation of results for the mass of the strange quark from
  the FLAG report\,\cite{Colangelo:2010et}. The top panel shows
  results from simulations with dynamical up, down and strange quarks
  ($\Nf=2+1$), while the middle panel refers to QCD with two light
  flavours. In the bottom panel (blue points) results from non-lattice
  determinations are plotted. Red symbols denote results that have
  been published in refereed journals. Results from preprints or
  conference proceedings are represented by green points. The global
  estimate for QCD with $\Nf=2+1$ flavours is represented by the grey
  band, while the corresponding estimate for $\Nf=2$ flavours is
  bounded by the dashed lines.}
\label{fig:mstrange}
\end{center}
\end{figure}

Surely, the r\^ole of lattice QCD is not restricted to the
verification of known results. For instance, the masses of the light
quarks are not directly accessible by experiment but can be predicted
by lattice calculations. Recent years have witnessed the publication
of a wealth of lattice results for the strange quark mass, as well as
the isospin-averaged light quark mass. The availability of accurate
predictions for a number of phenomenologically relevant observables
has prompted the foundation of the FLAG Working Group which tries to
form global averages for quark masses, meson form factors and decay
constants, and other quantities, very much in the spirit of the
Particle Data Group. Figure\,\ref{fig:mstrange} shows a compilation of
results for the mass of the strange quark from the FLAG
review\,\cite{Colangelo:2010et}. A remarkable feature is the
impressive level of consistency, despite the strong variation of
systematic effects among the different determinations. FLAG's analysis
of all published lattice data leads to the ``global'' estimates
of~\cite{Colangelo:2010et}
\be
   m_s^{\msbar}(2\,\GeV) = 94\pm3\,\MeV,\qquad 
   m_{ud}^{\msbar}(2\,\GeV) = 3.43\pm0.11\,\MeV,
\ee
where $m_{ud}=\half(m_u+m_d)$. Here the masses are quoted in the
$\msbar$-scheme of dimensional regularisation, at the commonly used
reference scale of\,$2\,\GeV$. This level of precision exceeds that of
the values quoted in the current edition of the Particle Data
Book\,\cite{PDG10} by an order of magnitude.

Among the other quantities discussed in the FLAG review are form
factors for $K_{\ell3}$ decays and the ratio $f_{\rm{K}}/f_\pi$ of
kaon and pion decay constants. When combined with the experimentally
measured branching fractions for the respective leptonic and
semi-leptonic decays, the lattice estimates can be used to test the
unitarity of the first row of the CKM matrix. As discussed in detail
in\,\cite{Colangelo:2010et}, the test can be further strengthened by
including the constraint of the direct determination of $|V_{ud}|$
from nuclear $\beta$-decay. In this way, first-row unitarity is
confirmed at the permille level, using only lattice results and
experimental data as input.

\begin{figure}
\begin{center}
\includegraphics[width=9cm]{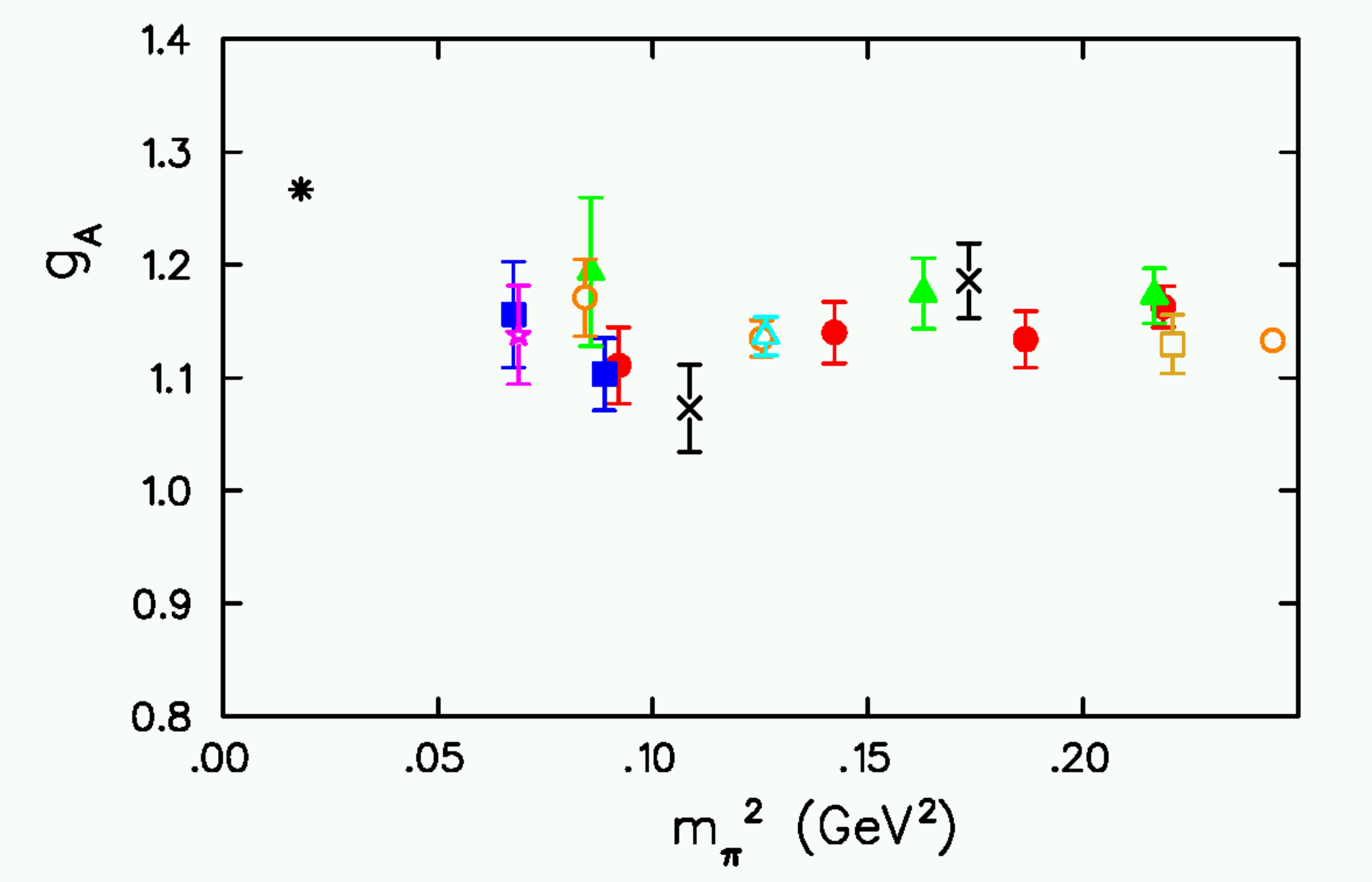}
\caption{Recent compilation of lattice results for the nucleon axial
  charge, plotted versus the pion mass squared (from
  ref.\,\cite{alexandrou_lat10}). The various sets of symbols refer to
  data from different collaborations. The left-most point denotes the
  value from the Particle Data Book.}
\label{fig:gAsumm}
\end{center}
\end{figure}

In spite of these successes, one finds examples for which the
agreement between lattice calculations and experiment is less
satisfactory. Observables that describe structural properties of the
nucleon fall into this category, as was pointed out in several recent
reviews on the
subject\,\cite{zanotti_lat08,renner_lat09,alexandrou_lat10}. For
instance, the experimentally observed dependence of the (isovector)
electric and magnetic form factors of the nucleon on the squared
momentum transfer $q^2$ is not reproduced. Moreover, lattice
calculations of the nucleon axial charge, $\gA$, lie typically
$10-15$\,\% below the experimental value, as signified by the summary
plot in Fig.\,\ref{fig:gAsumm}. What is even more disturbing is the
absence of a clear trend in the lattice data which would indicate that
the gap becomes narrower as the pion mass is decreased towards its
physical value. Since one would hesitate to conclude that QCD has been
falsified on the basis of these observations, the contradiction can
only be resolved if one accepts that systematic effects in lattice
calculations of these observables are not fully controlled.

The goal of our project is the calculation of a variety of nucleon and
meson form factors with complete control over all systematic
uncertainties. As we shall see, the application of novel techniques
which allow for a more reliable identification of the ground state is
a crucial ingredient for our task.

The outline for the remainder of this article is as follows: In
section~\ref{sec:s2latt} we review the basic ``lattice technology'',
including a discussion of the main systematic effects inherent in any
lattice calculation. Section~\ref{sec:s3pionff} describes our on-going
project aimed at producing a benchmark calculation of the pion
electromagnetic form factor. In section~\ref{sec:s4nuclff} we discuss
our determination of the nucleon form factors and the axial
charge. Finally, section~\ref{sec:s5concl} contains our conclusions.

\section{Lattice technology \label{sec:s2latt}}

Every lattice simulation proceeds by performing a stochastic
calculation of observables via Monte Carlo integration. If the gluon
field is represented by the link variables $U_\mu(x)$, the expectation
value of some observable $\Omega$ is given by
\be
   \left\langle\Omega\right\rangle = \frac{1}{Z}\int \prod_{x,\mu}
   {\rmd}U_\mu(x)\, \Omega \,\rme^{-S_{\rm G}[U]}\,\prod_f
    \det\left(\not\hspace{-3pt}{D}^{\rm lat}
    +m_f\right),
\ee
where ${\rmd}U_\mu(x)$ denotes an integration over the group manifold
of the gauge group SU(3). Each flavour $f=u,d,s,\ldots$ contributes a
factor of the quark determinant, where $\not\hspace{-3pt}{D}^{\rm
lat}$ is a suitable representation of the massless Dirac operator. It
is important to bear in mind that the discretisation is not unique:
Standard discretisations such as Wilson or staggered fermions have
been known for a long time. At the end of the 1990s, alternative
lattice transcriptions of the quark part of the action based on the
Ginsparg-Wilson relation\,\cite{GWR} were shown to preserve chiral
symmetry at non-zero lattice spacing\,\cite{HasLaNie,gwr:Luscher}, and
particular realisations such as domain
wall\,\cite{dwf:Kaplan,dwf:FurSha} or overlap
fermions\,\cite{ovlp:Neuberger} have been applied in practical
simulations. More recently, several different implementations of
so-called ``minimally doubled'' fermions have been
studied\,\cite{mind:Wilczek87,mind:Creutz07,mind:Borici07,mind:Creutz08,mind:Capitani09,mind:Capitani10},
in the hope of finding discretisations which have good chiral
properties but avoid the large inherent numerical costs of domain wall
or overlap quarks.

A common feature of all discretisations is the strong growth in the
computer time required for generating statistically independent
configurations of gauge fields as the light quark masses are tuned
towards the physical values of the up- and down-quark
masses\,\cite{panel_lat01}. During the past 10 years this algorithmic
problem has been greatly ameliorated owing to several significant
technical improvements. These include hierarchical integration
schemes\,\cite{Sexton:1992nu,DDHMC,Urbach05}, mass
preconditioning\,\cite{Hasenbus_trick}, domain decomposition
methods\,\cite{DDHMC}, deflation techniques\,\cite{DDHMC-defl} and
suitably optimised combinations thereof\,\cite{MPHMC_lat10}. The
value of the pion mass, i.e. the lightest mass in the pseudoscalar
meson channel, serves as a measure for how deeply a particular
simulation has penetrated into the chiral regime. As a result of the
recent algorithmic improvements one can now routinely access pion
masses as low as {200\,\MeV}, while in 2001 that figure stood at a
heavy 500\,{\MeV}. In a few cases, contact with the physical pion mass
has already been made\,\cite{dspect:PACS-CS09,quark:BMW10}.

We will now discuss the main systematic effects in lattice
calculations. For any non-zero value of the lattice spacing~$a$,
observables receive corrections of order~$a^p$, where the integer~$p$
depends on the chosen discretisation. Thus, the value of~$p$
determines the rate of convergence towards the continuum limit. In
practical simulations the continuum limit is taken by computing
observables for several values of~$a$ before performing an
extrapolation to $a=0$. The relatively high numerical cost of
simulations in the chiral regime implies that results at the physical
pion mass are obtained via chiral extrapolations. Although the chiral
behaviour of many observables can be constrained by Chiral
Perturbation Theory (ChPT), such extrapolations are still a major
source of uncertainty. This is particularly true for quantities which
describe structural properties of the nucleon.

Observables computed on the lattice are also affected by the finite
spatial volume. Empirically, one finds that uncertainties arising from
the finite box size~$L$ are subdominant, provided that $L$ is at least
as large as $2.5-3\,\fm$ and that the pion mass in units of~$L$
satisfies $m_{\pi}L\;\gtaeq\;3-4$. Whether this is true for any given
observable must be established on a case-by-case basis. In fact, there
are suspicions that the above bounds are not sufficient to guarantee
small finite-size effects for many baryonic
quantities\,\cite{renner_lat09}. It should also be noted that
finite-volume effects can be computed analytically in ChPT, which
offers the possibility of applying finite-volume corrections to the
final lattice estimates.

\begin{figure}
\begin{center}
\leavevmode
\includegraphics[width=6.3cm]{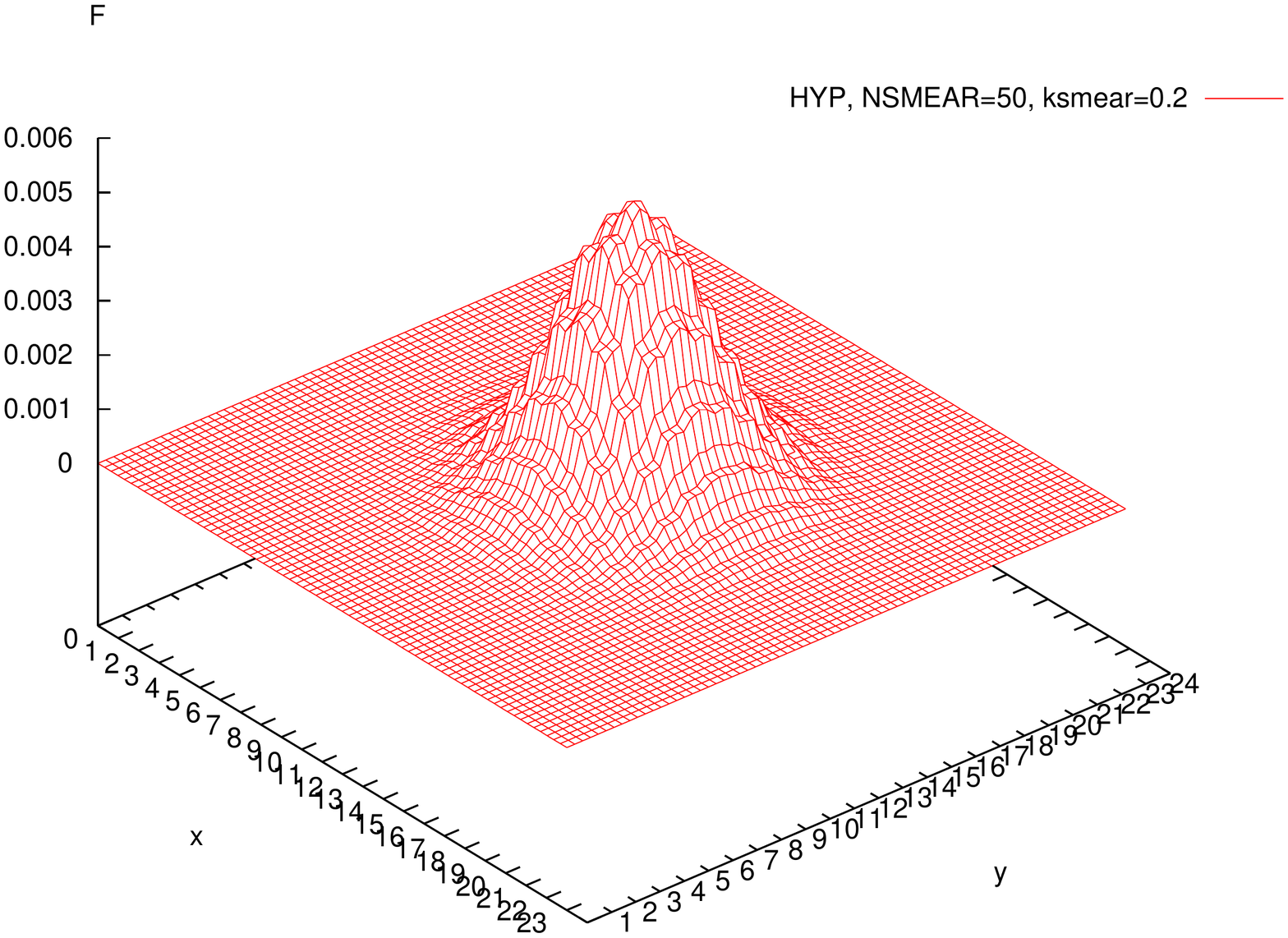}
\includegraphics[width=6.3cm]{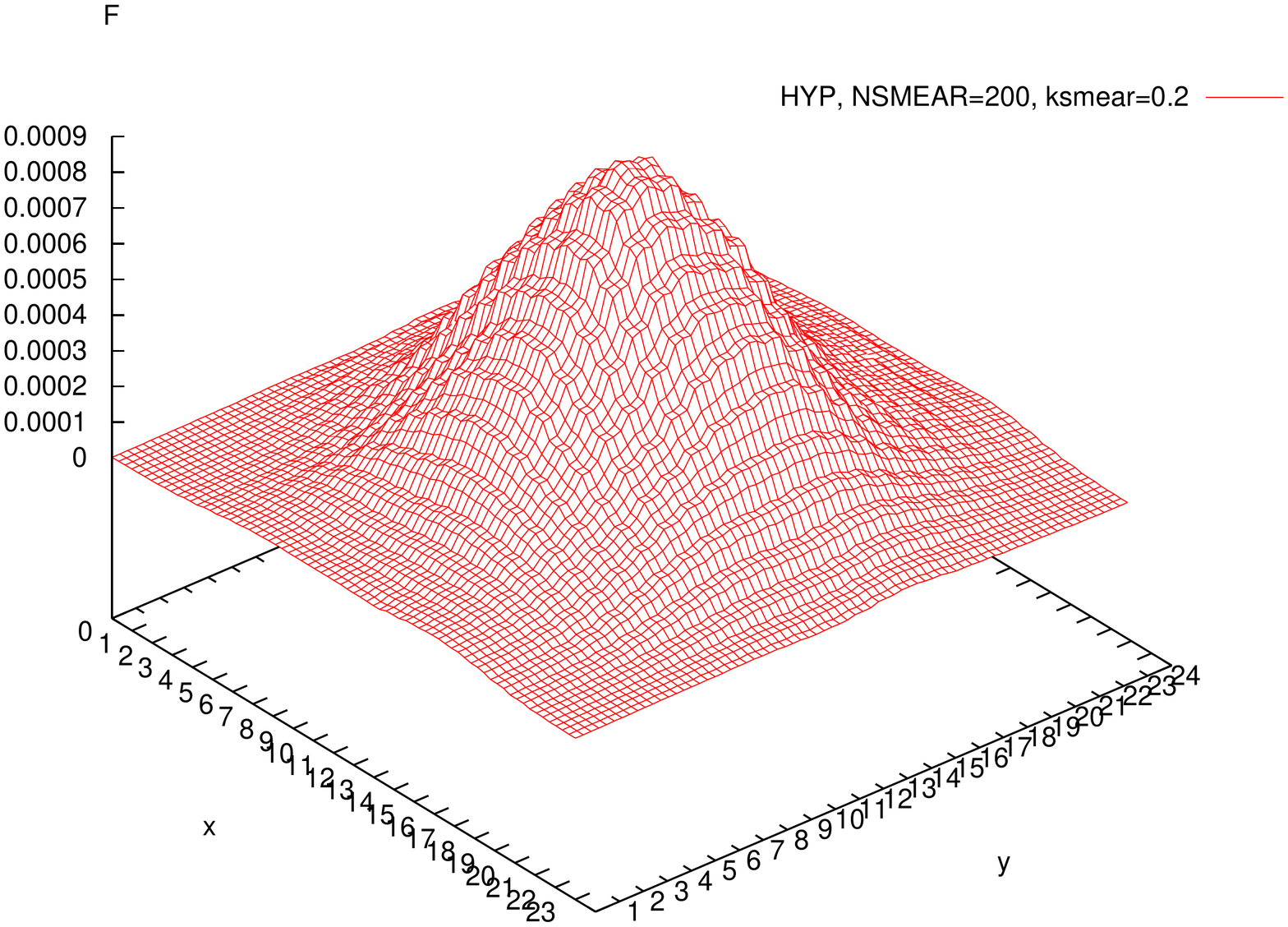}
\caption{\label{fig:smearing} Two spatial profiles of smeared quark
  fields obtained using the Jacobi algorithm\,\cite{smear:Jacobi93} in
  combination with HYP-smeared link variables\,\cite{smear:HYP01}. The
  width of the profile is controlled by the number of iterations and
  the value of the mass parameter in the kernel of the smearing
  function.}
\vspace{0.3cm}
\end{center}
\end{figure}

An important systematic effect arises from the possibility that masses
and matrix elements of ground state hadrons are contaminated by
contributions from excited states. Physical observables are usually
extracted from correlation functions of local composite fields. If
$O_{\rm{had}}$ denotes an interpolating operator for a particular
hadron, then its two-point correlation function is given by
\be
  \sum_{{\scriptsize\xvec,\yvec}}\rme^{i{\scriptsize\pvec}
    \cdot({\scriptsize\yvec-\xvec})} 
  \left\langle O_{\rm{had}}(y) O_{\rm{had}}^\dagger(x)\right\rangle
  =\sum_{n} w_n(\pvec) \rme^{-E_n({\scriptsize{\pvec}})(y_0-x_0)}
  \sim w_1(\pvec)\rme^{-E_1({\scriptsize{\pvec}})(y_0-x_0)},
\label{eq:twopt}
\ee
where the sum over~$n$ arises due to the insertion of a complete set
of eigenstates of the Hamiltonian. The quantity $w_n(\pvec)$ denotes
the spectral weight of the $n$th state. The operator $O_{\rm{had}}$
projects on all states that are characterised by the same quantum
numbers. The last relation shows that the ground state dominates the
correlation function for large Euclidean time separations,
$(y_0-x_0)\gg0$, which allows for the determination of the ground
state energy $E_1(\pvec)$ from the exponential fall-off. If, however,
the statistical fluctuations grow rapidly with increasing $(y_0-x_0)$
it may happen that the signal is lost before the contributions from
excited states are sufficiently suppressed. As a result, one incurs an
uncontrolled distortion of ground state properties. A widely used
procedure which is designed to enhance the projection onto the ground
state (i.e. which increases its weight $w_1(\pvec)$ in the spectral
representation) is called ``smearing''. Common smearing algorithms
apply a kernel function $F(\xvec,\yvec;U)$ to the quark field at point
$\yvec$, in order to approximate the spatial profile of the hadron's
wave function. Two examples of smeared sources are shown in
Fig.\,\ref{fig:smearing}

Our simulations have been performed on the high-performance cluster
``Wilson'' operated by the Institute of Nuclear Physics at Mainz
University. We use $\Nf=2$ flavours of $O(a)$ improved Wilson fermions
as our discretisation of the quark action. The non-perturbative
estimates for the improvement coefficient $\csw$ which multiplies the
Sheikholeslami-Wohlert term\,\cite{impr:SW} were taken from
ref.\,\cite{impr:csw_nf2}. Monte Carlo ensembles were generated for
three different values of the lattice spacing,
i.e. $a\approx0.08,\,0.07$ and $0.05\,\fm$ and for a range of pion
masses from $m_{\pi}\approx250\,\MeV$ to $700\,\MeV$. Lattice volumes
were chosen sufficiently large so as to satisfy $m_{\pi}L> 4$ on all
ensembles. A compilation of simulation parameters is shown in
Table\,\ref{tab:params}. The Monte Carlo ensembles are being generated
as part of the CLS (``Coordinated Lattice Simulations'')
project.\footnote{{\tt
https://twiki.cern.ch/twiki/bin/view/CLS/WebHome}}

\begin{table}
\begin{center}
\caption{\label{tab:params} Simulation parameters and approximate
  values for the lattice scale and pion masses. The preliminary
  results presented in this review are based on the ensembles labelled
  ``A'', ``E'', ``F'' and ``N''.}
\begin{tabular}{ccccccc}
\noalign{\vskip0.3ex}
\hline\hline\noalign{\vskip0.3ex}
 $\beta$ &    $a[\fm]$    & lattice   &  $L [\fm]$ & \# masses   &
 $m_{\pi}L$ & Labels \\
\noalign{\vskip0.3ex}
\hline\noalign{\vskip0.3ex}
  5.20 & 0.08 & $64 \times 32^3 $ & 2.6 &  4 masses & 4.8 -- 9.0 &
  $\sf A1 - A4$ \\ \noalign{\vskip0.3ex}
\hline\noalign{\vskip0.3ex}
  5.30 & 0.07 & $48 \times 24^3 $ & 1.7 &  3 masses & 4.6 -- 7.9 &
  $\sf D1 - D3$ \\ \noalign{\vskip0.3ex}
  5.30 & 0.07 & $64 \times 32^3 $ & 2.2 &  3 masses & 4.7 -- 7.9 &
  $\sf E3 - E5$ \\ \noalign{\vskip0.3ex}
  5.30 & 0.07 & $96 \times 48^3 $ & 3.4 &  2 masses & 5.0, 4.2   &
  $\sf F6, F7$ \\  \noalign{\vskip0.3ex}
\hline \noalign{\vskip0.3ex}
  5.50 & 0.05 & $96 \times 48^3 $ & 2.5 &  3 masses & 5.3 -- 7.7 &
  $\sf N3 - N5$ \\ \noalign{\vskip0.3ex}
  5.50 & 0.05 & $128\times 64^3 $ & 3.4 &  1 mass   & 4.7 & $\sf O7$
  \\ \noalign{\vskip0.3ex}
\hline\hline
\end{tabular}
\end{center}
\end{table}

\section{The pion electromagnetic form factor \label{sec:s3pionff}}

The electromagnetic form factor, defined by
\be
   \left\langle\pi^+(\pvec_f)|
   {\textstyle\frac{2}{3}}\ubar\gamma_\mu u
  -{\textstyle\frac{1}{3}}\dbar\gamma_\mu d
   |\pi^+(\pvec_i)\right\rangle = (p_f+p_i)_\mu\,f_\pi(q^2),
\ee
where $q=p_f-p_i$ is the momentum transfer, encodes the distribution
of electric charge inside the pion. Of particular interest is the
charge radius, $\langle r^2_\pi\rangle$, which is derived from the
pion form factor at vanishing momentum transfer, i.e.
\be
       f_\pi(q^2)=1-\frac{1}{6}\langle r_\pi^2\rangle
       q^2+\rmO(q^4) \quad\Rightarrow\quad
       \langle r^2_\pi\rangle = 6\left.
       \frac{\rmd f_\pi(q^2)}{\rmd q^2}\right|_{q^2=0}.
\ee
Lattice calculations of mesonic matrix elements are technically
simpler than the corresponding quantities for the
nucleon. Furthermore, the pion electromagnetic form factor receives no
contributions from quark-disconnected diagrams, whose evaluation
typically suffers from large statistical fluctuations. This opens the
possibility to perform a precision test of lattice QCD, by comparing
lattice estimates for $\langle r^2_\pi\rangle$ to the experimentally
determined value. However, owing to the finite spatial volume the
accessible range of momentum transfers $q^2$ is severely constrained,
which presents a major obstacle for precise lattice determinations of
$\langle r^2_\pi\rangle$.

\begin{figure}
\begin{center}
\includegraphics[width=5cm]{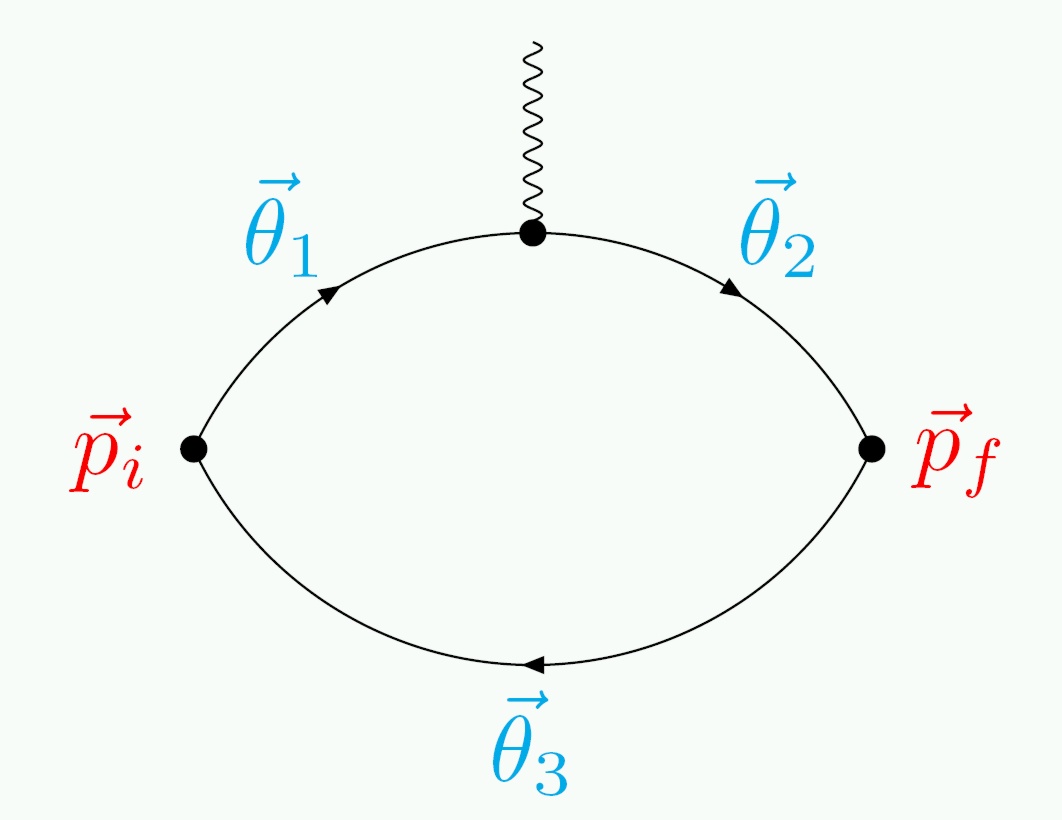}
\caption{\label{fig:3ptpion} Assignment of twist angles for the
  three-point correlation function of the vector current.}
\vspace{0.3cm}
\end{center}
\end{figure}

The solution to this problem, which by now has become a standard
method, is to employ partially twisted boundary
conditions\,\cite{twbc:Bedaque,twbc:deDi04}. To this end one imposes
periodicity on the quark fields up to a general phase, i.e.
\be
   \psi(x+L\mbf{\hat{e}^{(k)}}) = \rme^{i\theta^{(k)}} \psi(x), \qquad
   k=1,2,3, 
\ee
where $\mbf{\hat{e}^{(k)}}$ denotes a unit vector in the $k^{\rm{th}}$
spatial direction, and $\theta^{(k)}$ is the corresponding phase
angle.
A non-zero value of the latter modifies the accessible values of the
spatial momentum according to
\be
   \pvec={\mbf{n}}\frac{2\pi}{L}+\frac{{\mbf{\theta}}}{L}.
\ee
Quark propagators computed for different values of the twist angles
can be combined to form the three-point correlation functions from
which the pion form factor can be determined (see
Fig.\,\ref{fig:3ptpion}). The twist angles for the initial and final
state pions are then given by\,\cite{FFtwbc:UKQCD07}
\be
   {\mbf{\theta}}_i = {\mbf{\theta}}_1-{\mbf{\theta}}_3, \qquad
   {\mbf{\theta}}_f = {\mbf{\theta}}_2-{\mbf{\theta}}_3,
\ee
so that the expression for the squared momentum transfer becomes
\be
    -Q^2\equiv q^2=(p_f-p_i)^2 =
   \Big(E_\pi(\pvec_f)-E_\pi(\pvec_i)\Big)^2
  -\Big[\Big(\pvec_f+\frac{{\mbf{\theta}}_f}{L}\Big)
  -\Big(\pvec_i+\frac{{\mbf{\theta}}_i}{L}\Big) \Big]^2.
\ee
Thus, by an appropriate choice of twist angles one can tune $q^2$ to
any desired value. In our simulations we have chosen
${\mbf{\theta}}_i,\,{\mbf{\theta}}_f$ so as to achieve a particularly
fine momentum resolution near $q^2=0$.

So far we have assumed that boundary conditions are identical for sea
and valence quarks. It is, however, customary to apply
{\it{partially}} twisted boundary conditions, where the twist is
applied only to the quark fields in the valence sector. This has the
advantage that the generation of Monte Carlo ensembles must be
performed only once (e.g. for zero twist), while the choice of twist
angle and, in turn, the momentum transfer can be optimised for a
particular observable. The modification of the boundary conditions can
lead to finite-size effects associated with the breaking of flavour
symmetries. However, it was shown in ref.\,\cite{twbc:SachVilla04}
that such finite-size effects are exponentially suppressed in
processes without final-state interactions. Hence, for the
electromagnetic interaction between a photon and a pion finite-size
effects are expected to be small.

\begin{figure}
\begin{center}
\leavevmode
\includegraphics[height=4.8cm]{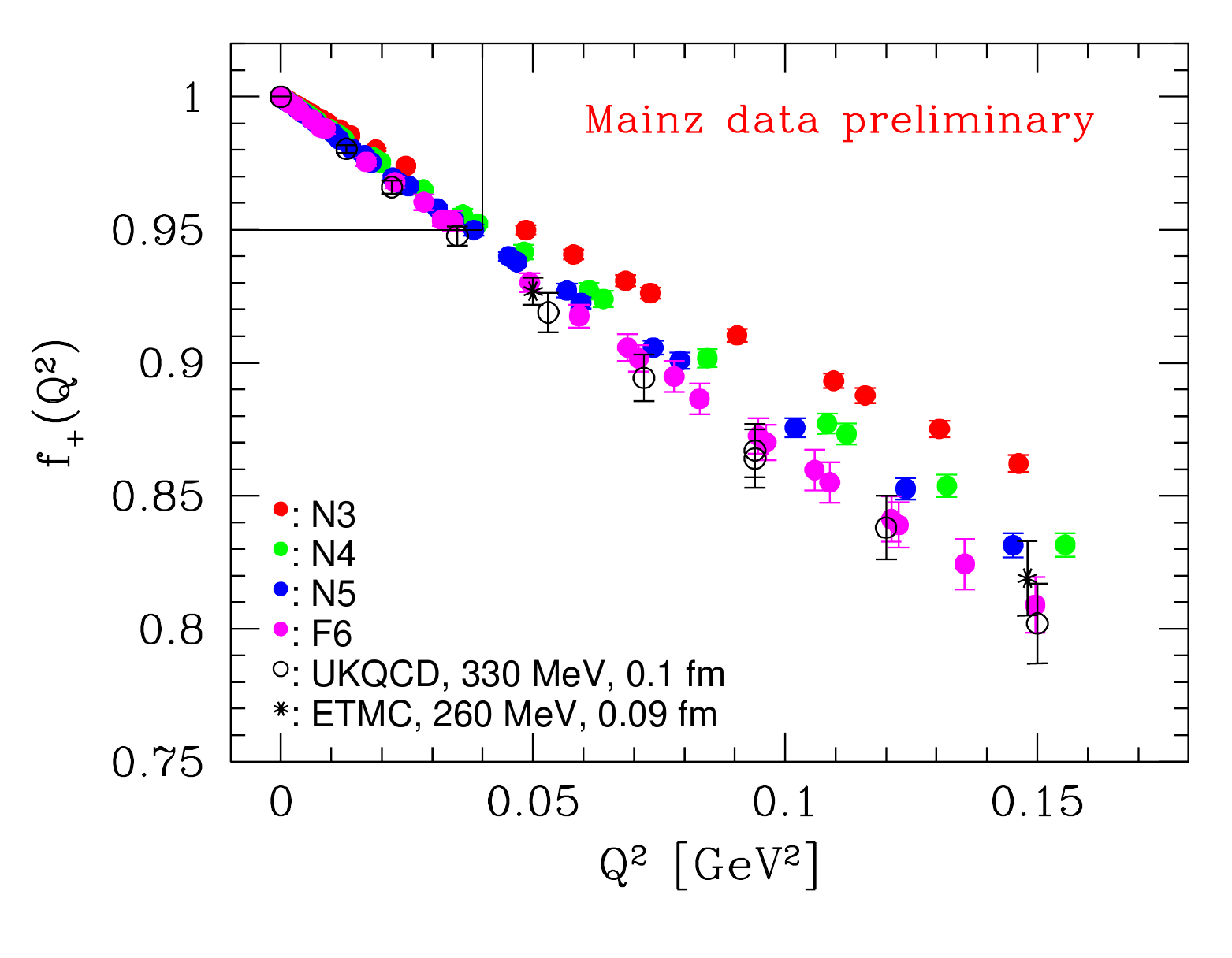}
\includegraphics[height=4.8cm]{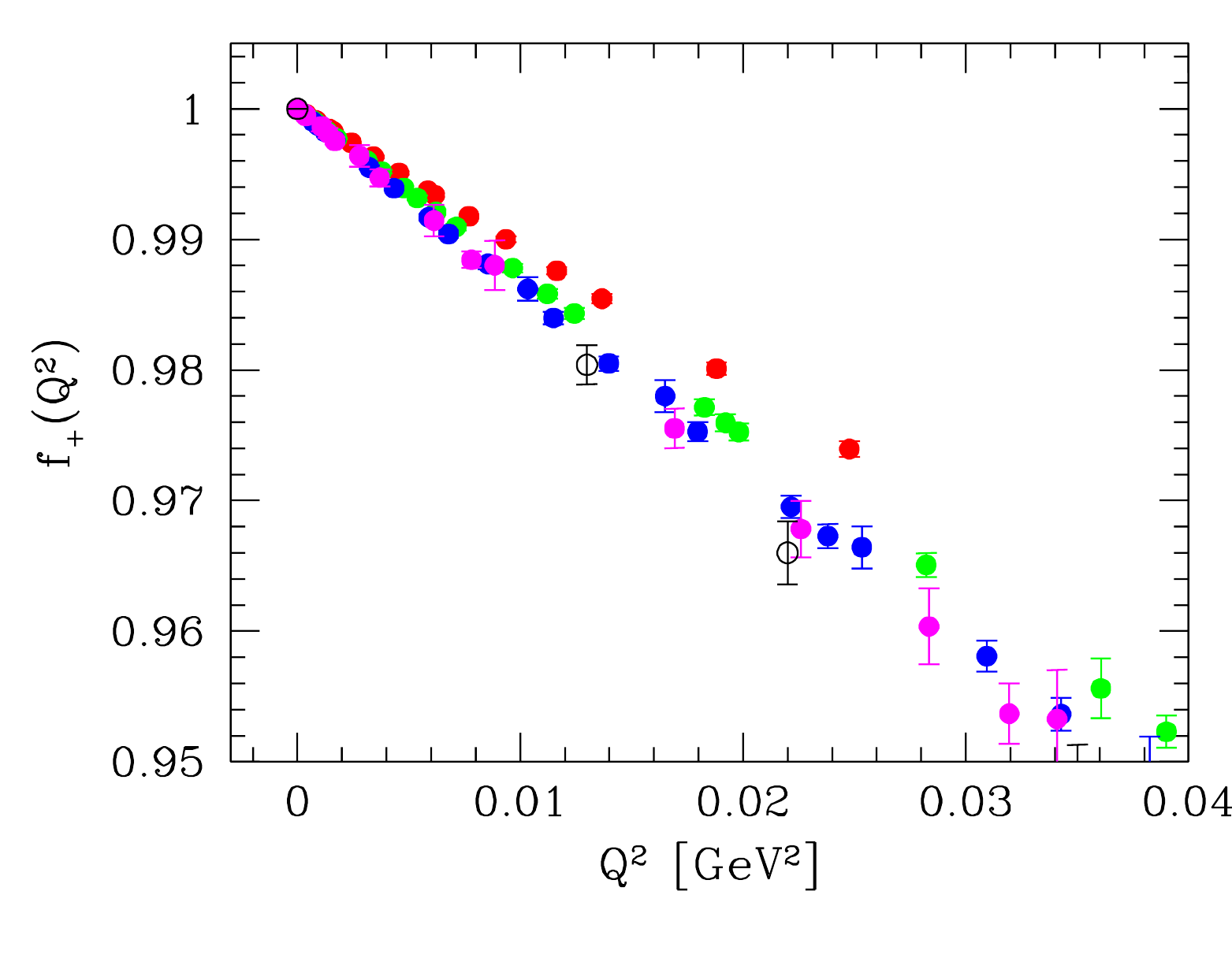}
\caption{\label{fig:pionff} {\bf Left:} pion form factor computed for
  a range of pion masses compared to the results of
  \cite{FFtwbc:UKQCD08,FFtwbc:ETM08}. {\bf Right:} data points from
  the inset in the top left-hand corner.}
\end{center}
\end{figure}

\begin{figure}
\begin{center}
\includegraphics[height=6.5cm]{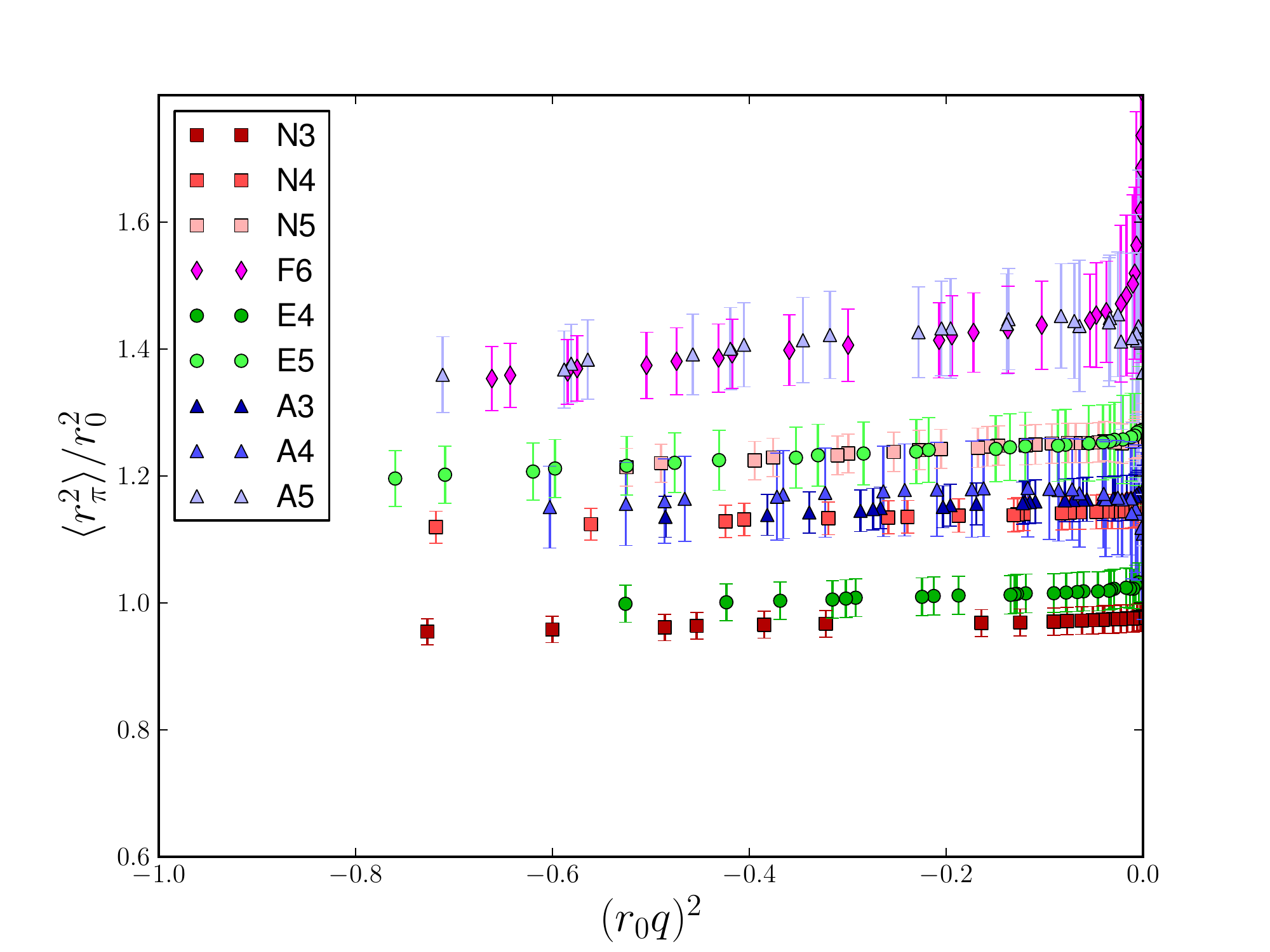}
\caption{\label{fig:rpisq} The squared pion charge radius (in units
  of $r_0$) extracted from the linear slope of the form factor in an
  interval $[0,(qr_0)^2]$, plotted versus the interval length. The
  meaning of the labels is given in Table\,\ref{tab:params}.}
\end{center}
\end{figure}

\begin{figure}
\begin{center}
\leavevmode
\includegraphics[height=7cm]{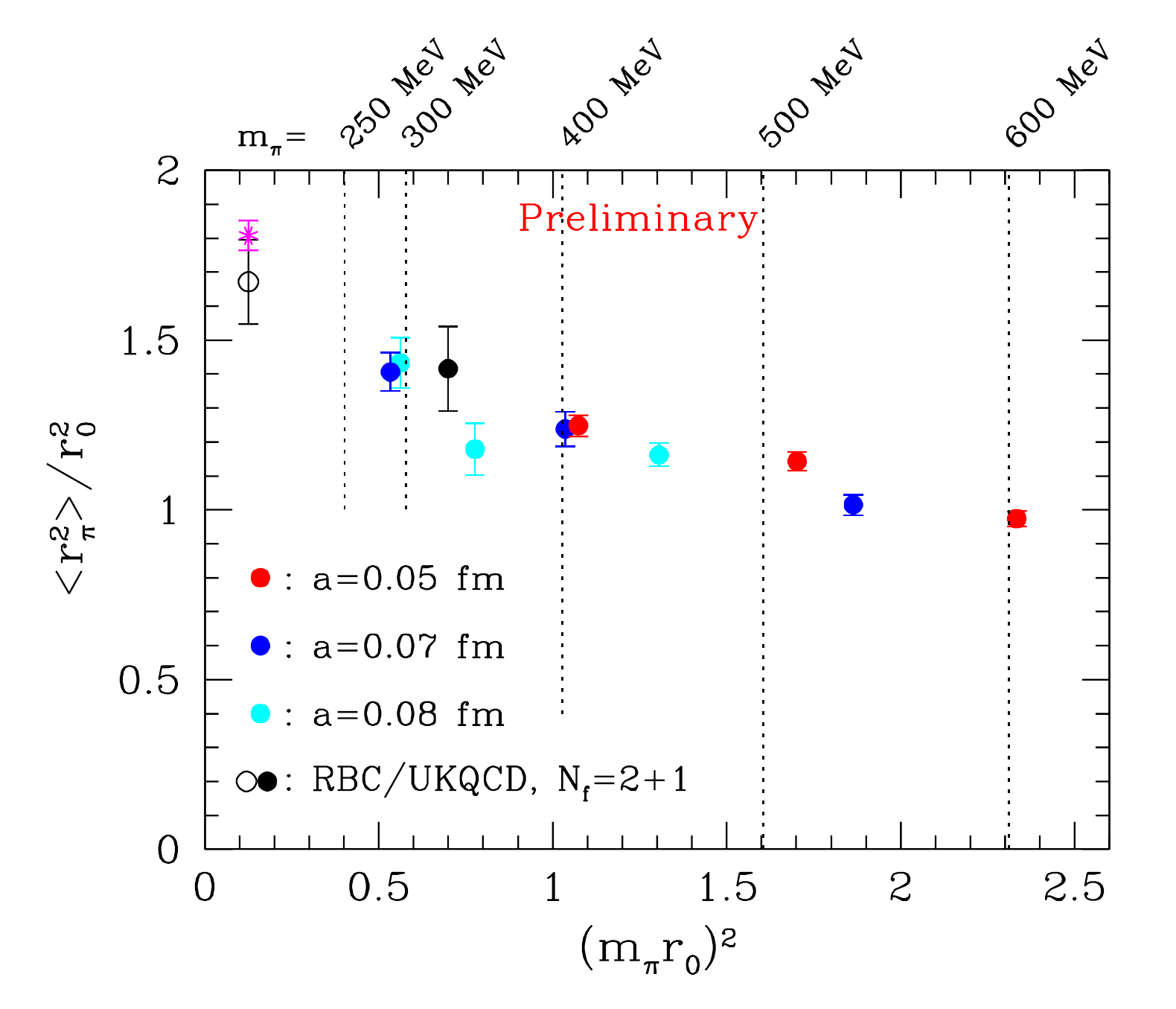}
\caption{\label{fig:chargeradius} The squared pion charge radius as a
  function of the squared pion mass. The black open and solid symbols
  are taken from ref.\,\cite{FFtwbc:UKQCD08}. The value from the
  Particle Data Book is indicated by the pink star. All dimensionful
  quantities are expressed in units of the hadronic radius $r_0$. To
  locate the positions of the pion masses in physical units we have
  set $r_0=0.5\,\fm$.}
\end{center}
\end{figure}

In Fig.\,\ref{fig:pionff} we show our results for the pion form factor
computed on the ensembles N3, N4, N5 and F6. There are two main
observations: First, there is a clear trend towards a steeper fall-off
with $q^2=-Q^2$ as the pion mass decreases from about 600\,{\MeV} on
N3 to about 290\,{\MeV} on F6. Secondly, by our choice of twist angles
we were able to produce a very dense set of points near $q^2=0$. This
allows us to extract the pion's charge radius in an accurate and
model-independent fashion, by determining the linear slope of
$f_{\pi}(q^2)$ over a narrow interval, starting at $q^2=0$. For the
following discussion we express all dimensionful quantities in units
of the hadronic radius $r_0$\,\cite{pot:r0,pot:r0_nf2}. In
Fig.\,\ref{fig:rpisq} the values of $\langle r_\pi^2\rangle/r_0^2$ are
plotted versus the length of the interval in $(qr_0)^2$ over which the
slope was determined. Obviously one would like to choose this interval
as small as possible. The figure shows that the statistical accuracy
in the determination of $\langle r_\pi^2\rangle/r_0^2$ is still very
good in the immediate vicinity of vanishing momentum transfer. The
fact that the resulting estimates of $\langle r_\pi^2\rangle$ are
practically constant implies that terms of $\rmO(q^4)$ in the chiral
expansion of $f_{\pi}$ are quite small.

Our preliminary results for the squared charge radii were determined
from the linear slope over the interval
$-0.15\leq(qr_0)^2\leq0$. Using $r_0=0.5\,\fm$ this corresponds to
$|q|\;\lesssim\;150\,\MeV$. Figure\,\ref{fig:chargeradius} shows the
chiral behaviour of the charge radius. By comparing the solid red and
blue points we conclude that our data are accurate enough to exhibit a
sensitivity to lattice artefacts. Overall, though, the chiral trend
compares favourably with the phenomenological value of $\langle
r_\pi^2\rangle$, shown as the pink symbol in the plot. A more
systematic investigation of lattice artefacts, as well as a more
detailed study of the $q^2$-dependence of the form factor is left for
future work.

\section{Nucleon form factors and axial charge} \label{sec:s4nuclff}

\begin{figure}
\begin{center}
\includegraphics[width=12cm]{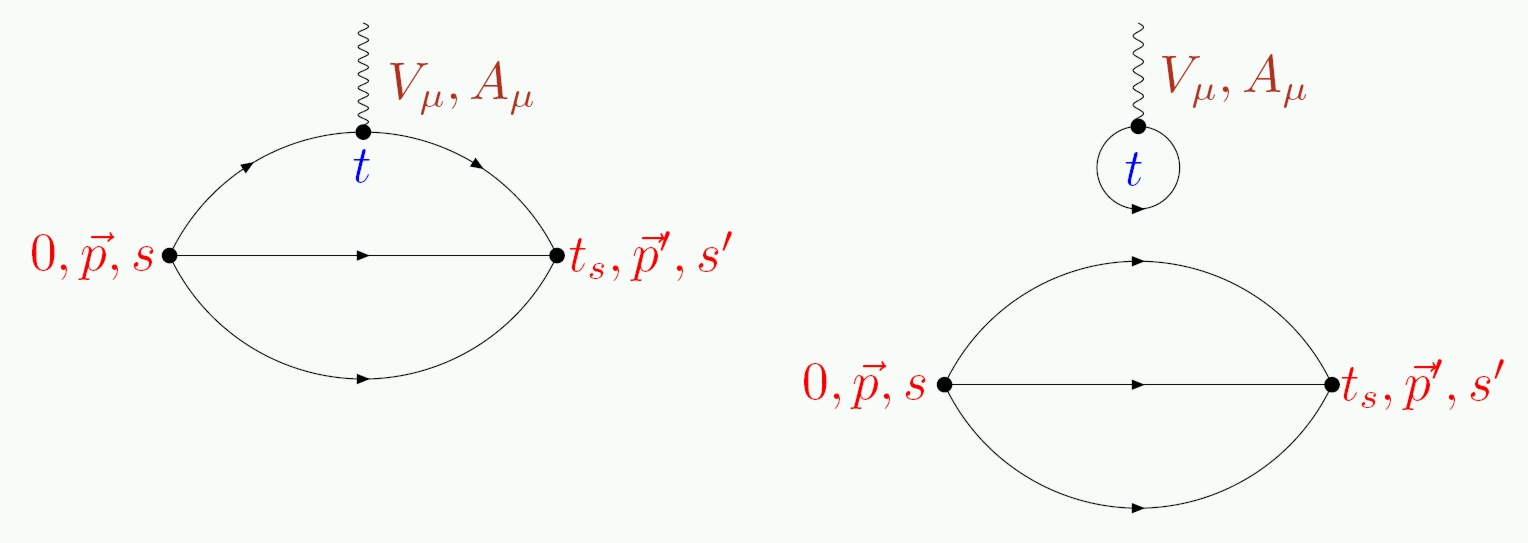}
\caption{\label{fig:diags} Quark-connected and -disconnected
  contributions to three-point correlation functions of the vector and
  axial-vector currents between nucleon states.}
\vspace{0.3cm}
\end{center}
\end{figure}

The pion form factor discussed in the previous section can be
considered a warm-up exercise for the technically more difficult case
of extracting the corresponding quantities in the nucleon sector. The
well-known Dirac and Pauli form factors, $F_1$ and $F_2$, are related
to the matrix element of the electromagnetic vector current between
nucleon initial and final states via
\be
   \left\langle N(p',s')\left| V_\mu(x)
      \right|N(p,s)\right\rangle =\ubar(p',s')\Big[ \gamma_\mu
      F_1(q^2) +\sigma_{\mu\nu}\frac{q_\nu}{2m_{\rm N}} F_2(q^2)
      \Big] u(p,s),
\ee
where $\left|N(p,s)\right\rangle$ denotes the initial state of a
nucleon with momentum~$p$ and spin~$s$. The electric and magnetic form
factors $G_{\rm{E}}$ and $G_{\rm{M}}$ are derived from
\be
   G_{\rm E}(q^2) = F_1(q^2) -\frac{q^2}{(2m_{\rm N})^2} F_2(q^2),\quad
   G_{\rm M}(q^2) = F_1(q^2)+ F_2(q^2).
\ee
Similarly, the matrix element of the axial current $A_\mu^a(x)\equiv
\psibar(x)\gamma_\mu\gamma_5\half\tau^a\psi(x)$ is parameterised in
terms of the form factors $G_{\rm{A}}$ and $G_{\rm{P}}$, i.e.
\be
   \left\langle N(p',s')\left| A_\mu^a(x)
      \right|N(p,s)\right\rangle =\ubar(p',s')\half\tau^a\Big[
        \gamma_\mu\gamma_5 G_{\rm{A}}(q^2) 
       + \frac{q_\mu\gamma_5}{2m_{\rm N}} G_{\rm{P}}(q^2)
      \Big] u(p,s).
\ee
The axial charge, $\gA$, is defined as the axial form factor at
vanishing momentum transfer,
$\displaystyle\gA=\lim_{q^2\to0}G_{\rm{A}}(q^2)$. Nucleon matrix
elements are determined by computing the corresponding three-point
correlation functions. After performing the Wick contractions one can
express the correlators in terms of quark propagators. In general,
this gives rise to the quark-connected and quark-disconnected diagrams
depicted in Fig.\,\ref{fig:diags}. Since the disconnected contribution
is statistically very noisy, it is mostly neglected in lattice
calculations, which is the approach we have adopted as well during the
first stage of this project.

Unlike the case of the pion form factor there have been only a few attempts to
apply partially twisted boundary conditions to calculations of nucleon form
factors. This more cautious approach is motivated by the observation that the
induced finite-volume effects can become sizeable for small twist angles
\cite{Jiang:2008ja}. 

It was already mentioned in the introduction that lattice calculations
have so far failed to reproduce the experimentally observed dependence
of electromagnetic form factors on the momentum transfer. This, in
turn, implies that the associated charge radii are not consistent
either. Moreover, lattice calculations tend to underestimate the axial
charge $\gA$ by $10-15$\,\%. Uncontrolled systematic effects are held
responsible for this. With only a few exceptions, nucleon form factors
have mainly been computed over a very limited range of lattice
spacings\,\cite{nuclFF:LHPC02_nf2,nuclFF:LHPC05_nf2p1,nuclFF:QCDSF06_nf2,nuclFF:RBC08_nf2,nuclFF:RBC08_nf2p1,nuclFF:RBC09_nf2p1,nuclFF:LHPC09_nf2p1,nuclFF:LHPC10_nf2p1,nuclFF:QCDSF_lat10,nuclff:QCDSF_lat10gA,nuclff:ETMC1011_nf2}. It
is thus conceivable that the agreement between lattice calculations
and experiment is improved when performing a systematic continuum
extrapolation to $a=0$. Also, nucleon form factors could be more
sensitive to finite-volume effects, compared to their mesonic
counterparts\,\cite{renner_lat09}. Another potentially very important
source of systematic error is the chiral extrapolation. The quark
masses used in current simulations are still relatively large and
leave a long extrapolation to the physical value of the
(isospin-averaged) light quark mass. In particular, the region around
the physical pion mass is not well constrained by the data, and the
functional form for the extrapolation may be too crude to yield
reliable results.

What makes the calculation of nucleon form factors a much more
difficult task compared to the corresponding mesonic quantities, is
the relatively high inherent level of statistical noise in baryonic
correlation functions. While the signal-to-noise ratio of pseudoscalar
meson correlators can be shown to remain constant as the Euclidean
time separation is increased, one finds that the noise increases
exponentially in the baryonic
sector\,\cite{Parisi:1983ae,Thacker:1990bm,Luscher:2010ae}. The high
level of statistical fluctuations is the source of another potentially
very dangerous systematic effect, namely the contamination of ground
state properties by contributions from excited states, which have not
died out in the region of Euclidean times where statistical errors are
still quite small. Unless one has interpolating operators at one's
disposal which maximise the spectral weight of the ground state (see
\eq{eq:twopt}) one either risks the distortion of results from excited
state contributions or has to accept large statistical errors,
provided that the signal is not lost entirely. The commonly used
smearing methods to enhance the spectral weight of the ground state
may be insufficient to guarantee reliable results. They must be
combined with more sophisticated techniques such as the generalised
eigenvalue problem\,\cite{gevp:CMI83,phaseshifts:LW,gevp:alpha09}, or,
as in our project, summed operator
insertions\,\cite{Maiani:1987by,Gusken:1989ad}.

In order to illustrate our approach we restrict ourselves to the
discussion of the nucleon axial charge. For several reasons, $\gA$ is
an ideal benchmark quantity for lattice calculations of structural
properties of the nucleon: (1) $\gA$ is derived from a matrix element
of a simple quark bilinear which contains no derivatives, (2) initial
and final nucleon states are at rest, and (3) its definition as an
iso-vector quantity implies that contributions from quark-disconnected
diagrams are absent.

Nucleon form factors are usually extracted from suitably chosen ratios
of three- and two-point functions, such as
\be
    R_\Gamma({\mbf{q}};t,t_s)=
    \frac{C_3^\Gamma({\mbf{q}},t,t_s)}{C_2({\mbf{0}},t_s)} \cdot 
\left\{ \frac{C_2({\mbf{q}},t_s-t)\,C_2({\mbf{0}},t)\,C_2({\mbf{0}},t_s)}
             {C_2({\mbf{0}},t_s-t)\,C_2({\mbf{q}},t)\,C_2({\mbf{q}},t_s)}
\right\}^{1/2},
\ee
where $\Gamma=V,\,A$ characterises the current which is inserted at
time~$t$. The nucleon is created from the vacuum at time zero and
annihilated at time $t_s>t$. For $\gA$, the momentum transfer
$q=p^\prime-p$ is zero. Assuming that the axial current is correctly
renormalised and that all kinematical factors are properly taken into
account the ratio $R_{\rm{A}}$ gives direct access to $\gA$, i.e.
\be
   R(t,t_s)\equiv R_{\rm{A}}({\mbf{q}}=0,t,t_s) = \gA
   +\rmO(\rme^{-{\Delta}t}) +\rmO(\rme^{-{\Delta}(t_s-t)}). 
\label{eq:Rconv}
\ee
Here, $\Delta$ denotes the mass gap between the ground state nucleon
and its first excitation. In QCD with dynamical quarks $\Delta$ is
expected to be equal to $2m_\pi$, which implies that the corrections
to $\gA$ in \eq{eq:Rconv} can be rather sizeable, unless large values
of~$t$ and~$t_s$ are considered. However, since the statistical errors
grow exponentially with time separation, it is difficult to optimise
the choice of~$t$ and~$t_s$, in order to keep both the systematic
errors due to excited state contamination and the statistical errors
under control. Most published results for $\gA$ were obtained by
fitting the ratio $R(t,t_s)$ to a constant in~$t$, for
$t_s\approx1\,\fm$. Below we present evidence that this procedure may
be insufficient to ensure that the resulting estimates of $\gA$ are
free from excited state contributions.

Let $C_3^{\rm{A}}(t,t_s)$ denote the three-point correlation function
of the axial current for ${\mbf{q}}=0$. For a fixed value of $t_s$ we
define the ratio $\rho(t,t_s)$ as
\be
  \rho(t,t_s):=\frac{C_3^{\rm{A}}(t,t_s)}{C_3^{\rm{A}}(t_s/2,t_s)},
  \qquad t=a, 2a,\ldots,t_s.
\ee
Obviously, the deviation of $\rho(t;t_{\rm s})$ from unity at a
particular value of~$t$ indicates the presence of excited state
contributions. If the three-point functions are computed using smeared
sources (and perhaps also sinks) the deviation of $\rho(t,t_s)$ from
one can be regarded as a measure of the effectiveness of the smearing
procedure. Since the axial charge is usually determined by fitting
$R(t,t_s)$ to a constant for $t$ around $t_s/2$, the parameters of the
smearing procedure must be tuned such that excited state contributions
are eliminated inside the fit range.

\begin{figure}
\begin{center}
\includegraphics[width=9.5cm]{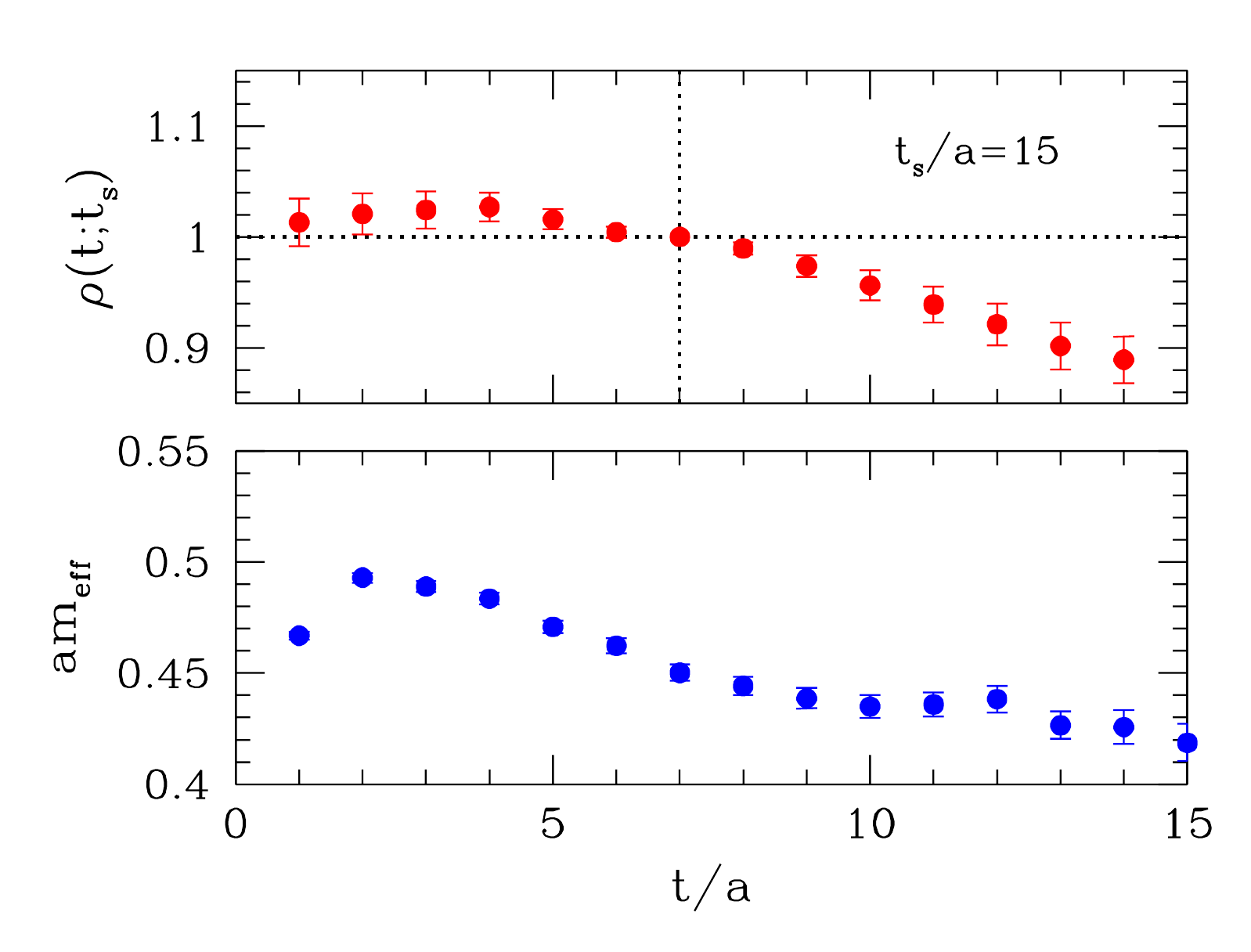}
\caption{The ratio $\rho(t,t_s)$ (top panel) and the effective mass of
  the nucleon (bottom panel), computed on a lattice of size
  $32^3\cdot64$ ($L=2.2\,\fm$) and a pion mass of 415\,\MeV. Two- and
  three-point functions were computed using Jacobi-smeared
  sources. The parameters in the smearing procedure were chosen to
  produce a smearing radius of $\approx\,0.5\,\fm$.}
\label{fig:rhoSL}
\end{center}
\end{figure}

In Fig.\,\ref{fig:rhoSL} the ratio $\rho(t,t_s)$ computed using
smeared-local (SL) correlators\footnote{The notation ``SL'' is used to
describe a correlator which is smeared at the source, $t=0$, only.}
for $t_s=1.1\,\fm$ is plotted against~$t$. While $\rho$ is mostly
compatible with one for small values of $t$, there are large and
significant deviations for $t>t_{\rm s}/2$. One concludes that source
smearing is unable to remove excited state contamination in the
interval $0\;{\leq}\;t\;{\leq}\;t_s$. The problem is further
highlighted by comparing $\rho(t,t_s)$ to the effective mass of the
nucleon: The bottom panel in Fig.\,\ref{fig:rhoSL} shows clearly that
the asymptotic behaviour of the nucleon two-point function only sets
in at timeslices $t$ for which the deviation of $\rho$ from unity
becomes significant. We can draw the conclusion that source smearing
alone cannot guarantee the absence of contamination from excited
states in ratios such as $R(t,t_{\rm{s}})$. Our findings suggest that
there is a mismatch in the asymptotic behaviour of three- and
two-point functions which enter the ratio $R(t,t_s)$. As a
consequence, the appearance of a plateau in $R(t,t_s)$ for
$t,\,t_s\;\lesssim\;1\,\fm$ must considered to be mostly
accidental. One might expect that the situation improves if
correlators are smeared both at the source, $t=0$, and sink,
$t=t_s$. This is currently under investigation.

The use of summed operator insertions\,\cite{Maiani:1987by} offers a
handle for eliminating excited state contributions. The key
observation is that the corrections from excited states in
\eq{eq:Rconv} can be parametrically reduced. Restricting the
discussion once more to the case of the axial charge, one defines the
summed ratio $S(t_s)$ via
\be
   S(t_s) := \sum_{t=0}^{t_s}\,R(t,t_s).
\ee
Its asymptotic behaviour is given by
\be
   S(t_s) = c +{\gA}t_s +\rmO(t_s\rme^{-{\Delta}t_s}),
\ee
where the (divergent) constant~$c$ includes contributions from contact
terms. Since $t_s>t$ the corrections to $S(t_s)$ are in general more
strongly suppressed than for $R(t,t_s)$. The reduction of excited
state contributions in $S(t_s)$ comes at a price, though, since the
summed ratio must be computed for several different values of $t_s$,
in order to extract $\gA$ from the linear slope. As a further comment
we add that the method can be straightforwardly extended to cases
where the initial and final states have different momenta. The general
expression is given as
\be
   S_{\Gamma}({\mbf{q}};t_s) :=
   \sum_{t=0}^{t_s}\,R_{\Gamma}({\mbf{q}};t,t_s) = K +M(q^2)t_s+
   \rmO(t_s\rme^{-{\Delta}t_s}) +\rmO(t_s\rme^{-{\Delta^\prime}t_s}), 
\ee
where $M(q^2)$ is the matrix element of interest, and $\Delta,
\Delta^\prime$ denote the energy gaps in the initial and final states,
respectively. 

\begin{figure}
\begin{center}
\includegraphics[width=9.0cm]{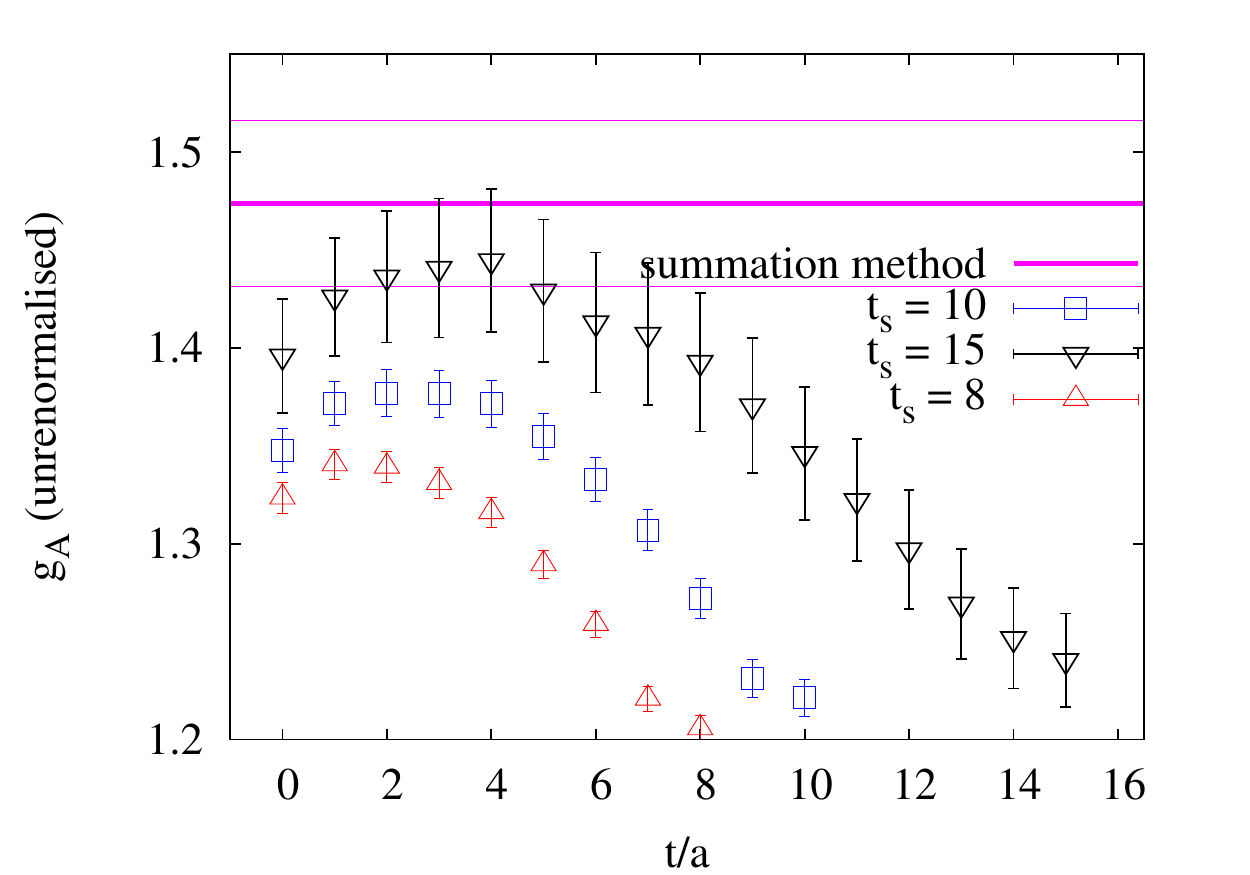}
\caption{The ratio $R(t,t_s)$ of the iso-vector axial charge computed
  for three different values of $t_s$ at $\beta=5.3$
  ($a\approx0.07\,\fm$). The purple band denotes the result extracted
  from the slope of the summed correlator.} 
\label{fig:gAmethods}
\end{center}
\end{figure}

In our simulations we have computed the summed correlator ratios for
$t_s\approx 0.7 - 1.1\,\fm$, using both the vector and axial vector
currents. As an illustration how the method works, we compare in
Fig.\,\ref{fig:gAmethods} the $t$-dependence of the ratio $R(t,t_s)$
computed for three different values of $t_s$ against the result
extracted from the linear slope of the summed correlator. The latter
is shown as the purple band in the figure. It is seen that
$t_s=15a\approx 1.1\,\fm$ is just sufficient to produce a plateau
which agrees with the result from the summed insertion within
statistical errors. Nonetheless, it is clear that the summed correlator
produces a larger value for $\gA$. 

The case for using summed correlators is even more compelling as the
chiral limit is approached. Since $\Delta\approx 2m_\pi$ decreases for
lighter quark masses, the correction terms in eq.\,(\ref{eq:Rconv})
become larger and may spoil the expected chiral behaviour of $\gA$.
The size of the corrections in the standard approach also depends on
the spatial extent $L$ of the lattice, since the overlap of a local
interpolating field with a multi-particle state is suppressed by
powers of the volume. In the conventional approach it is then
difficult to disentangle finite-volume effects from excited states
contaminations. A slight drawback of the method can be read off from
Fig.\,\ref{fig:gAmethods}: Excited state contributions to the summed
correlator are reduced at the expense of incurring larger statistical
errors.

\begin{figure}
\begin{center}
\includegraphics[width=9.0cm]{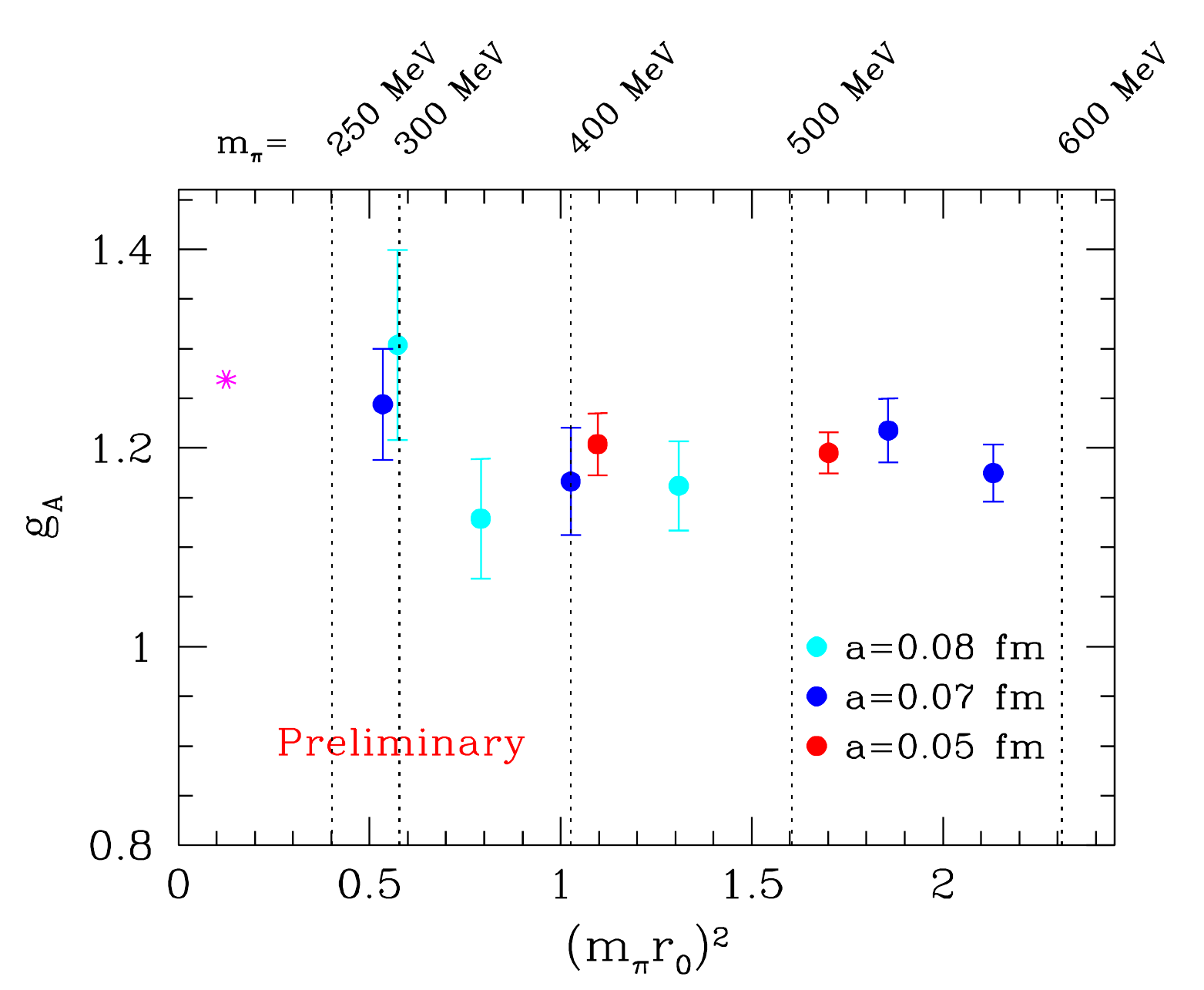}
\caption{The axial charge $\gA$ determined using summed correlators
  plotted versus the pion mass squared. The magenta star represents
  the PDG value of $\gA=1.2695(29)$.} 
\label{fig:gAresults}
\end{center}
\end{figure}

The current status of our $\gA$ determination is summarised in
Fig.\,\ref{fig:gAresults}. All data points were determined via the
slope of the summed correlators which results in larger statistical
errors compared to conventional calculations using similar
statistics. Nonetheless, it is clear that these preliminary results
are in good agreement with the experimental value for pion masses
$m_\pi\;\lesssim\;300\,\MeV$. These findings differ from what is
usually observed by other collaborations.
%
%
Clearly, further studies including data produced at smaller
pion masses and additional sets of correlation functions which are
smeared both at the source and the sink are required before any
definite conclusions can be drawn. Also, we will perform a systematic
investigation of the influence of lattice artefacts, by analysing the
results obtained at three values of the lattice spacing.

\section{Summary and conclusions} \label{sec:s5concl}

Despite the recent successes of lattice QCD it is clear that
quantities describing structural properties of hadrons are still
afflicted with one or several sources of systematic errors. In this
work we have outlined and applied a number of technical
improvements. In particular, summed insertions have proven a valuable
tool to suppress excited state contamination in ratios of correlation
functions. Moreover, partially twisted boundary conditions lead to a
much enhanced momentum resolution in calculations of the pion form
factor, which greatly facilitates the extraction of the pion's charge
radius with reduced model dependence. Finally, our large lattice
volumes and fine lattice spacings ensure that the corresponding
systematic effects are under good control.

Owing to its computational simplicity, the pion form factor is an
ideal testbed for the more complicated case of nucleon form
factors. Clearly, more work, including the calculation of correlators
which are smeared both at the source and sink, is required before any
definite statement about the computed $q^2$-dependence and its
comparison with experimental data can be made. The axial charge of the
nucleon, $\gA$, is of particular importance, since its determination
does not involve the technically challenging calculation of
disconnected diagrams, and because it is defined at a single value of
$q^2$. Both of these features make it an ideal reference quantity for
future benchmark calculations of structural properties of the nucleon.

\subsection*{Acknowledgments}

We thank our colleagues within the CLS project for sharing gauge
ensembles. Calculations of correlation functions were performed on the
dedicated QCD platform ``Wilson'' at the Institute for Nuclear
Physics, University of Mainz. This work is supported by DFG (SFB443),
GSI, and the Research Center~EMG funded by {\sl Forschungsinitiative
Rheinland-Pfalz}.


\end{document}